\documentclass{article}
\usepackage{spconf,amsmath, graphicx}
\usepackage{comment}
\usepackage{cite}
\usepackage{color}
\usepackage{enumerate}
\usepackage{array}
\usepackage{breqn}
\usepackage{amssymb}

\usepackage{multirow}
\usepackage{arydshln}
\title{Representation Learning For Speech Recognition Using Feedback Based Relevance Weighting}
\name{Purvi Agrawal and Sriram Ganapathy
 \thanks{This work was partly funded by grants from the Department of Atomic Energy (DAE/34/20/12/2018-BRNS/34088) project, and the Ministry of Human Resource and Development (MHRD), Government of India.}
}
\address{Learning and Extraction of Acoustic Patterns (LEAP) lab,\\ Electrical Engineering, Indian Institute of Science, Bangalore, India.}

\ninept
\begin{document}
\ninept
\maketitle
\begin{abstract}
  In this work, we propose an acoustic embedding based approach for representation learning in speech recognition. The proposed approach involves two stages comprising of acoustic filterbank learning from raw waveform, followed by modulation filterbank learning. In each stage, a relevance weighting operation is employed that acts as a feature selection module. In particular, the relevance weighting network receives embeddings of the model outputs from the previous time instants as feedback. The proposed relevance weighting scheme allows the respective feature representations to be adaptively selected before propagation to the higher layers. The application of the proposed approach for the task of speech recognition on Aurora-4 and CHiME-3 datasets gives significant performance improvements over baseline systems on raw waveform signal as well as those based on mel representations (average relative improvement of $15$\% over the mel baseline on Aurora-4 dataset and $7$\% on CHiME-3 dataset).
\end{abstract}

\section{Introduction}
\label{sec:intro}
Representation learning deals with the broad set of methods that enable the learning of meaningful representations from raw data. Similar to machine learning, representation learning can be carried out in an unsupervised fashion like principal component analysis (PCA), t-stochastic neighborhood embeddings (tSNE) proposed by \cite{maaten2008visualizing} or in supervised fashion like linear discriminant analysis (LDA). Recently, deep learning based  representation learning has drawn substantial interest. While a lot of success has been reported for text and image domains (for eg., word2vec embeddings \cite{mikolov2013efficient}), representation learning for speech and audio is still challenging. 

One of the research directions pursued for speech has been the learning of filter banks operating directly on the raw waveform \cite{doss2013, sainath2013, tuske2014acoustic, hoshen2015speech, sainath2015cldnn}, mostly in supervised setting. 
Other efforts attempting unsupervised learning of filterbank have also been investigated. The work in \cite{sailor2016filterbank} used restricted Boltzmann machine while the efforts in \cite{agrawal2019unsupervised} used variational autoencoders. The wav2vec method recently proposed by \cite{schneider2019wav2vec} explores unsupervised pre-training for speech recognition by learning representations of raw audio.
There has been some attempts to explore interpretability of acoustic filterbank recently, for eg. SincNet filterbank by \cite{ravanelli2018interpretable} and self-supervised learning by \cite{pascual2019pase}. However, compared to vector representations of text which have shown to embed meaningful semantic properties, the interpretability of speech representations from these approaches has often been limited. 

Subsequent to acoustic filterbank processing, modulation filtering is the process of filtering the 2-D spectrogram-like representation using 2-D filters along the time (rate filtering) and frequency (scale filtering) dimension. Several attempts have been made to learn the modulation filters also from data. The earliest approaches using LDA explored the learning of the
temporal modulation filters in a supervised manner \cite{vuurenLda1997, hung2006optimization}. Using deep learning, there have been recent attempts to learn modulation filters in an unsupervised manner \cite{sailor2016unsupervised, agrawal2019jstsp}. 

In this paper, we extend our previous work \cite{agrawal2020interpretable} on joint acoustic and modulation filter learning in the first two layers of a convolutional neural network (CNN) operating on raw speech waveform. The novel contribution of our approach is the incorporation of acoustic embeddings as feedback in the relevance weighting approach. In particular, the relevance weighting network is driven by the acoustic/modulation filter outputs along with the embedding of the previous one-hot targets. 
The output of the relevance network is a relevance weight which multiplies the acoustic/modulation filter \cite{agrawal2020interpretable}.  The rest of the architecture performs the task of acoustic modeling for automatic speech recognition (ASR). 
The approach of feeding the model outputs back to the neural network is also previously reported as a form of recurrent neural network (RNN) called the teacher forcing network \cite{williams1989learning}.  
However, in this work, the embeddings of the model outputs are fed back only to the relevance weighting network and not as a RNN architecture.

The ASR experiments are conducted on Aurora-4 (additive noise with channel artifact) dataset \cite{hirsch2000aurora}, CHiME-3 (additive noise with reverberation) dataset \cite{barker2015chime3} and VOiCES (additive noise with reverberation) dataset \cite{nandwana2019voices}.
The experiments show that the learned representations from the proposed  framework 
provide considerable improvements in ASR results over the baseline methods.

\section{Relevance Based Representation Learning}
\begin{center}
    \begin{figure}[t]
      \hspace{-0.2cm}
        \includegraphics[trim={0.4in 3.3in 1.5in 1.0in}, clip, scale=0.56]{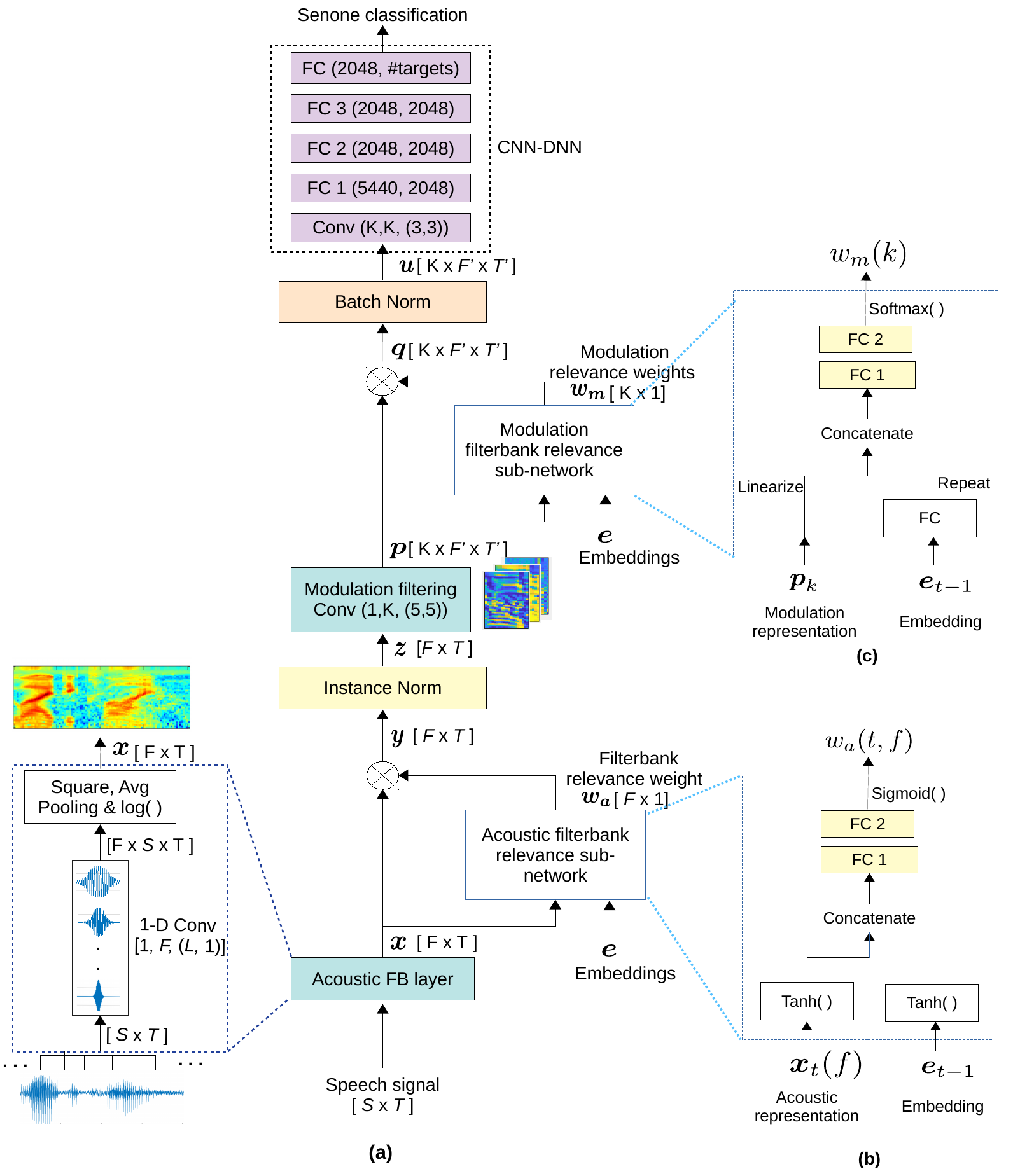}
        \vspace{-0.6cm}
        \caption{(a) Block diagram of the proposed representation learning approach from raw waveform, (b) expanded acoustic FB relevance sub-network. Here, $\boldsymbol{x}_t(f)$ denotes the sub-band trajectory of band $f$ for all frames centered at time $t$, $\boldsymbol{e}_{t-1}$ denotes the acoustic embedding vector for previous time step, (c) expanded modulation filterbank relevance sub-network.}
        \label{fig:block_diag}
        \vspace{-0.4cm}
    \end{figure}
\end{center}

{\label{sec:rep_learning}}
The block schematic of the senone embedding network is shown in Figure~\ref{fig:block_diag_embedding}. The entire acoustic model using the proposed relevance weighting model is shown in Figure~\ref{fig:block_diag}.

\subsection{Step-0: Embedding network pre-training}
\begin{center}
    \begin{figure}[t]
        \centering
        
        \includegraphics[trim={0 1.4in 0.2 0.05in}, clip, scale=0.19]{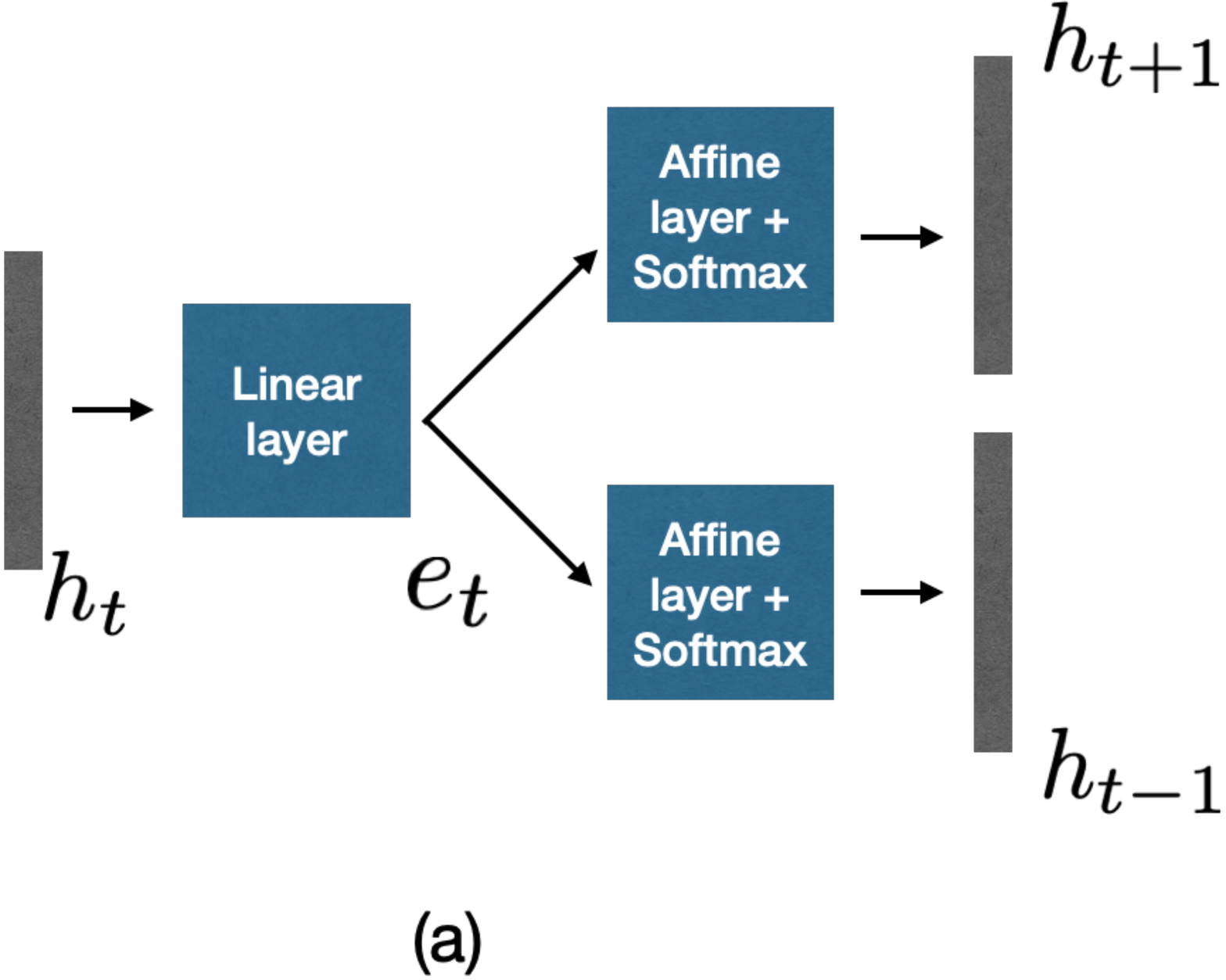}
        \vspace{-0.1cm}
        \caption{Block schematic of senone embedding network used in the proposed model.}
        \label{fig:block_diag_embedding}
    \end{figure}
\end{center}

The embedding network (Figure~\ref{fig:block_diag_embedding})  is similar to the skip-gram network of word2vec models as proposed in  \cite{mikolov2013efficient}. In this work, the one-hot encoded senone (context dependent triphone hidden Markov model (HMM) states modeled in ASR
) target vector at frame $t$, denoted as $\boldsymbol{h}_t$, is fed to a network whose first layer outputs the embedding denoted as $\boldsymbol{e}_t$. This embedding predicts the one-hot target vectors for the preceding and succeeding time frames $\boldsymbol{h}_{t-1}$ and $\boldsymbol{h}_{t+1}$. This model is trained using the ASR labels for each task before the acoustic model training. Once the model is trained, only the embedding extraction part (first layer outputs) is used in the final ASR model. We use embeddings of $200$ dimensions. During the ASR testing, the embeddings are derived by feeding the softmax outputs from the acoustic model (similar to teacher forcing network by \cite{williams1989learning}).

\begin{figure}[t!]
        \centering
       \includegraphics[trim={0.7cm 0 0 0.26in}, clip, scale=0.39]{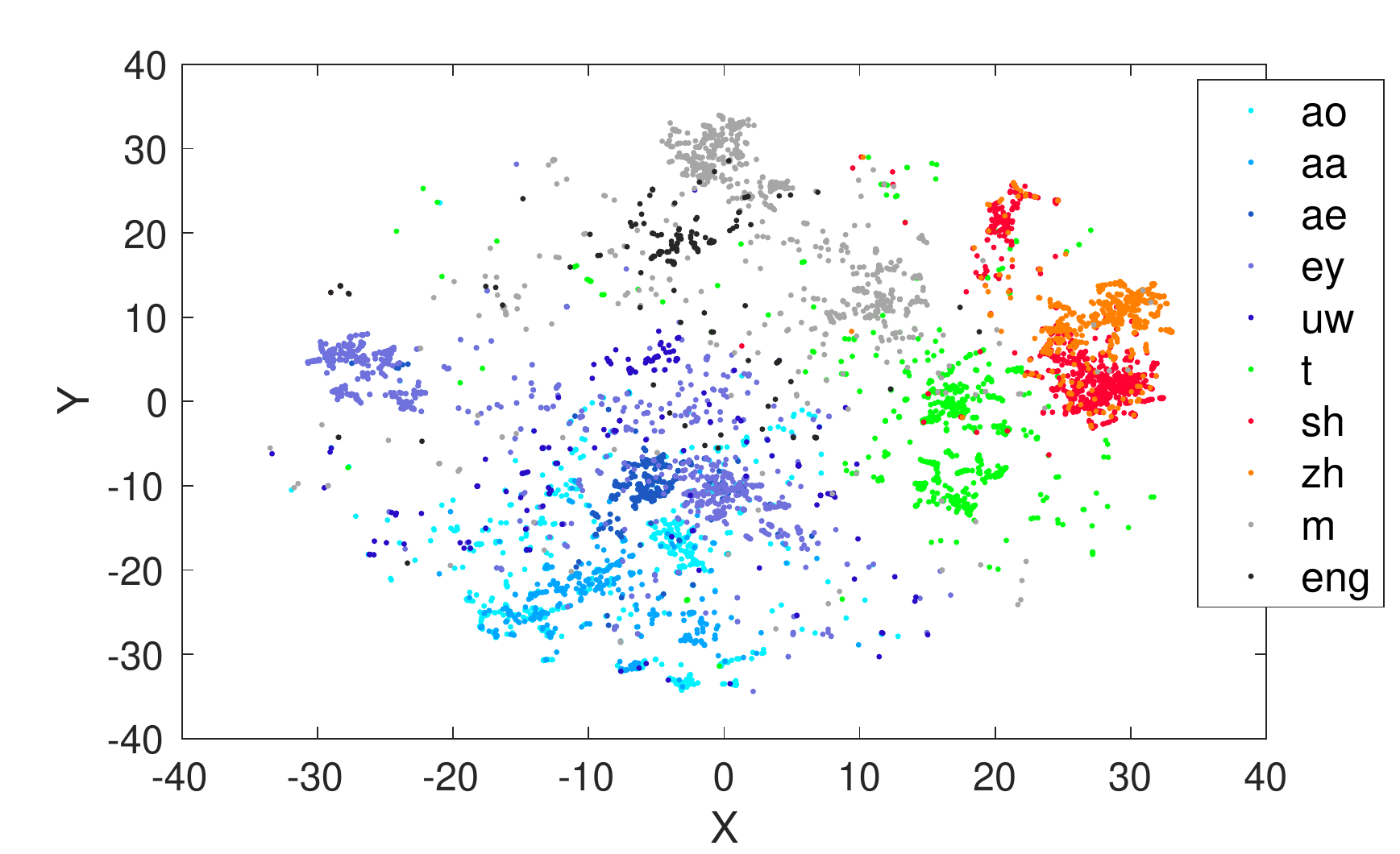}
        \vspace{-0.2cm}
        \caption{t-SNE plot of the senone embeddings for TIMIT dataset.}
        \label{fig:tsne_embedding_phonemes}
        \vspace{-0.4cm}
    \end{figure}

For the analysis, the TIMIT  test set \cite{timit1993darpa} consisting of $1344$ utterances is used. The dataset is hand labelled for phonemes. The t-SNE visualization of the embeddings is shown in Fig. \ref{fig:tsne_embedding_phonemes} for phonemes from TIMIT test set for a group of vowel phonemes \{/ao/, /aa/, /ae/, /ey/, /uw/\} and a group of plosives \{/t/\}, fricatives \{/sh/, /zh/\}, and nasals \{/em/, /eng/\}. As seen in the t-SNE plot of embeddings, the embeddings while being trained on one-hot senones, provides  segregation of different phoneme types such as vowels, nasals, fricatives and plosives.  

\vspace{-2.2cm}
\subsection{Step-1: Acoustic Filterbank representation \cite{agrawal2020raw}} 
 The input to the neural network are raw samples windowed into $S$ samples per frame with a contextual window of $T$ frames. Each block of $S$ samples is referred to as a frame. This input of size $S \times 1$ raw audio samples are processed with a 1-D convolution  using $F$ kernels ($F$ denotes the number of sub-bands in filterbank decomposition) each of size $L$.  The kernels are modeled as cosine-modulated Gaussian function \cite{agrawal2019unsupervised, agrawal2020raw},
\begin{equation}
    {{g}}_i (n) = \cos{2\pi\mu_i n} \times \exp{(-{n^2}\mu_i^2/{2})}
\end{equation}
where ${{g}}_i (n)$ is the $i$-th kernel ($i=1,..,F$)  at time $n$, $\mu_i$ is the center frequency of the $i$th filter (in frequency domain).
The mean parameter ${\mu}_i$ is updated in a supervised manner for each dataset.
The convolution with the cosine-modulated Gaussian filters generates $F$ feature maps which are squared, average pooled within each frame and log transformed. This generates $\boldsymbol{x}$ as $F$ dimensional features for each of the $T$ contextual frames, as shown in Figure \ref{fig:block_diag}. The $\boldsymbol{x}$ can be interpreted as the ``learned'' time-frequency representation (spectrogram). 

\subsection{Acoustic FB relevance weighting} {\label{subsec:self_relevance weighting}}
The relevance weighting paradigm for acoustic FB layer is implemented using a relevance sub-network fed with the $F \times T$ time-frequency representation $\boldsymbol{x}$ and embeddings $\boldsymbol{e}$ of the previous time step. Let $\boldsymbol{x}_t(f)$ denote the vector containing sub-band trajectory of band $f$ for all $T$ frames centered at $t$ (shown in Figure \ref{fig:block_diag}(b)). Then, $\boldsymbol{x}_t(f)$ is concatenated with embeddings of the previous time step $\boldsymbol{e}_{t-1}$ with $tanh()$ non-linearity. This is fed to a two layer deep neural network (DNN) with a sigmoid non-linearity at the output. It generates a scalar relevance weight $\boldsymbol{w}_a(t,f)$ as the relevance weight corresponding to the input representation at time $t$ for sub-band $f$.  This operation is repeated for all the $F$ sub-bands  which gives a $F$ dimensional weight vector $\boldsymbol{w}_a(t)$ for the input $\boldsymbol{x}_t$.

The $F$ dimensional weights $\boldsymbol{w}_a(t)$ multiply each column of the ``learned'' spectrogram representation $\boldsymbol{x}_t$ to obtain the relevance weighted filterbank representation $\boldsymbol{y}_t$. The relevance weights in the proposed framework are different from typical attention mechanism \cite{zhang2017attention}. In the proposed framework, relevance weighting is applied on the representation as soft feature selection weights without performing a linear combination. We also process the first layer outputs ($\boldsymbol{y}$) using instance norm \cite{rumelhart1986learning, ulyanov2016instance}. 

In our experiments, we use $T=101$ whose center frame is the senone target for the acoustic model. We also use $F=80$ sub-bands and acoustic filter length $L=129$. This value of $L$ corresponds to $8$ ms in time for a $16$ kHz sampled signal. The value of $S$ is $400$ ($25$ ms window length) with frame shift of $10$ms. 


\subsection{Step-2: Relevance Weighting of Modulation Filtered Representation} \vspace{-0.1cm}
The representation $\boldsymbol{z}$ from acoustic filterbank layer is fed to the second convolutional layer which is interpreted as modulation filtering layer (shown in Figure \ref{fig:block_diag}). The kernels of this convolutional layer are 2-D spectro-temporal modulation filters, learning the rate-scale characteristics from the data. 
The modulation filtering layer generates $K$ parallel streams, corresponding to $K$ modulation filters $\boldsymbol{w}_K$. The modulation filtered representations $\boldsymbol{p}$ are max-pooled with window of $3 \times 1$, leading to feature maps of size $F' \times T'$. These are weighted using a second relevance weighting sub-network (referred to as the modulation filter relevance sub-network in Figure \ref{fig:block_diag}, expanded in Figure \ref{fig:block_diag}(c)). 

The modulation relevance sub-network is fed with feature map $\boldsymbol{p}_k$; where $k=1,2,...,K$, and embeddings $e$ of the previous time step. The embedding $\boldsymbol{e}$ is linear transformed and concatenated with the input feature map. This is fed to a two-layer DNN with softmax non-linearity at the output. It generates a scalar relevance weight $w_m(k)$ corresponding to the input representation at time $t$ ($t$ as center frame) for $k$th feature map.
The weights $\boldsymbol{w}_m$ are multiplied with the representation $\boldsymbol{p}$ to obtain weighted representation $\boldsymbol{q}$.  
The resultant weighted representation $\boldsymbol{q}$ is fed to the batch normalization layer \cite{batch2015norm}.
We use the value of $K=40$ in the work. Following the acoustic filterbank layer and the modulation filtering layer (including the relevance sub-networks), the acoustic model consists of series of CNN and DNN layers with sigmoid nonlinearity. 
\vspace{-0.1cm}
\section{Experiments and Results}{\label{sec:experiments}}
The speech recognition system is trained using PyTorch \cite{paszke2017pytorch} while the Kaldi toolkit \cite{povey2011kaldi} is used for decoding and language modeling. 
The models are discriminatively trained using the training data with cross entropy loss and Adam optimizer \cite{kingma2014adam}. A hidden Markov model - Gaussian mixture model (HMM-GMM) system is used to generate the senone alignments for training the CNN-DNN based model.
The ASR results are reported with a tri-gram language model or using a recurrent neural network language model (RNN-LM).

For each dataset, we compare the ASR performance of the proposed approach of learning acoustic representation from raw waveform with acoustic FB (A) with relevance weighting (A-R) and modulation FB (M) with relevance weighting (M-R) denoted as (A-R,M-R),
traditional log mel filterbank energy (MFB) features (80 dimension), power normalized filterbank energy (PFB) features \cite{kim2012pncc}, 
mean Hilbert envelope (MHE) features \cite{mhec2015}, and excitation based (EB) features \cite{drugman2015robust}. We also compare performance with the SincNet method proposed in \cite{ravanelli2018interpretable}. 
Note that the modulation filtering layer (M) is part of the baseline model, and hence notation M is not explicitly mentioned in the discussion.
The neural network architecture shown in Figure \ref{fig:block_diag} (except for the acoustic filterbank layer, the acoustic FB relevance sub-network and modulation filter relevance  sub-network) is used for all the baseline features. 

\begin{center}
    \begin{table}[t]
        \centering
        \vspace{-0.2cm}
        \caption{Word error rate (\%) for different configurations of the proposed model for the ASR task on Aurora-4 dataset.
        }
        \label{tab:variants_ASR}
        \resizebox{\linewidth}{!}{
        \begin{tabular}{l|c|c|c|c|c}
        \hline
         \multirow{2}{*}{\textbf{Features}} & \multicolumn{5}{c}{\textbf{ASR (WER in \%)}} \\ \cline{2-6}
          & A & B & C & D & Avg.  \\\hline
           Baseline Raw Waveform (A,M) & 4.1 & 6.8 & 7.3 & 16.2 & 10.7 \\ \hline \hline  
          \multicolumn{6}{c}{Acoustic Relevance} \\ \cline{1-6}
   
         A-R,M [Softmax, no embedding] \cite{agrawal2020interpretable} & 3.6 & {6.4} & {8.1} & {15.1} & {10.0}\\
         
         A-R,M [Sigmoid, no embedding] & 3.4 & {6.4} & {6.7} & {15.5} & {9.9}\\
         A-R,M [Sigmoid, with senone embedding] & 3.4 & {6.2} & {6.7} & {14.5} & \textbf{{9.6}}\\ \hline  \hline 
        \multicolumn{6}{c}{Acoustic Relevance \& Mod. Relevance} \\ \cline{1-6} 
         A-R,M-R [Softmax, no embedding] \cite{agrawal2020interpretable} & 3.6 & {6.1} & {\textbf{6.0}} & {14.8} & {9.6}\\
         A-R,M-R [Sigmoid, no embedding] & 3.4 & {6.0} & {6.5} & {14.5} & {9.5}\\
         
         A-R,M-R [Sigmoid, with senone embeddings] & {\textbf{3.0}} & {\textbf{5.8}} & 6.2 &  \textbf{14.4} &  \textbf{9.1}\\
         \hline
        \end{tabular}
        }
        \vspace{-0.2cm}
    \end{table}
\end{center}

\begin{center}   
     \begin{table}[t]
            \centering
            \begin{center}
            \caption{Word error rate (\%) in Aurora-4 database with various feature extraction schemes with decoding using trigram LM (and RNN-LM in paranthesis).
            }
            \label{tab:multiData_aurora4}
            \resizebox{\linewidth}{!}{
            \begin{tabular}{l|c|c|c|c|c|c|c|c}
            \hline
            \textbf{Cond} & MFB & PFB  & {MHE} & ~EB~ & Sinc &  MFB-R & S-R,M-R & A-R,M-R\\ \hline
            \multicolumn{9}{c}{A. Clean with same Mic} \\ \hline
            Clean & 4.2 & 4.0 & 3.8 & 3.7 & 4.0 & 3.9 & 3.8 & \textbf{3.0 (2.9)}\\ \hline
            \multicolumn{9}{c}{B: Noisy with same Mic} \\ \hline
            Airport & 6.8 & 7.1 & 7.3 & - & 6.9 &  6.7 & 6.2 & \textbf{5.7} \\
            Babble & 6.6 & 7.4 & 7.4 & -  & 6.7 & 6.5 & 6.1 & \textbf{5.7} \\
            Car & {4.0} & 4.5 & 4.3 & - & 4.0 & 4.1 & 3.9 & \textbf{3.6} \\
            Rest. & 9.4 & 9.6 & 9.1 & - & 9.4 & 9.6 & 8.4 & \textbf{7.0} \\
            Street & 8.1 & 8.1 & 7.6 & - & 8.4 & 8.4 & 7.5 & \textbf{6.3} \\
            Train & 8.4 & 8.6 & 8.6 & - & 8.3 & 8.2 & 7.4 & \textbf{6.8} \\\hdashline
            Avg. & 7.2 & 7.5 & 7.4 & 6.0 & 7.3 & 7.2 & 6.6 & \textbf{5.8 (5.3)}\\
             \hline
            \multicolumn{9}{c}{C: Clean with diff. Mic}\\ \hline
		    Clean & 7.2 & 7.3 & 7.3 & \textbf{5.0} & 7.3 &  {7.1} & 6.8 & {6.2 (5.9)} \\ \hline
            \multicolumn{9}{c}{D: Noisy with diff. Mic}  \\ \hline
            Airport & 16.3 & 18.0 & 17.6 & - & 16.2 &  16.2 & \textbf{13.9} & {14.0} \\
            Babble & 16.7 & 18.9 & 18.6 & - & 17.6 &  16.9 & 16.0 & \textbf{15.0}\\
            Car & {8.6} & 11.2 & 9.6 & - &  9.0 & 8.9 & 7.9 & \textbf{8.0}\\
            Rest. & 18.8 & 21.0 & 20.1 & - & 19.0 & 18.8 & 19.2 & \textbf{18.5} \\
            Street & 17.3 & 19.5 & 18.8 & - & 17.3 & 17.8 & 16.6 & \textbf{15.8}\\
            Train\ & 17.6 & 18.8 & 18.7 & - & 18.1 & 17.9 & 16.6 & \textbf{{15.3}} \\ \hdashline
            Avg. & 15.9 & 17.9 & 17.3 & 15.8 & 16.2 & 16.1 & 15.1 & \textbf{14.4 (13.7)} \\
             \hline
            \multicolumn{9}{c}{Avg. of all conditions}  \\ \hline
            Avg. & 10.7 & 11.7 & {11.4} & 9.9 & 10.8 & 10.8 & 9.9 & \textbf{9.1 (8.7)}\\ \hline
            \end{tabular}
            }
        \end{center}
         \vspace{-0.3cm}
      \end{table}
  \end{center}
\vspace{-1.5cm}
\subsection{Aurora-4 ASR}
This database consists of  read speech recordings of $5000$ words corpus, recorded under clean and noisy conditions (street, train, car, babble, restaurant, and airport) at $10-20$ dB SNR. The training data has $7138$ multi condition recordings ($84$ speakers) with total $15$ hours of training data. The validation data has $1206$ recordings for multi condition setup. The test data has $330$ recordings ($8$ speakers) for each of the $14$ clean and noise conditions. The test data are classified into group A - clean data, B - noisy data, C - clean data with channel distortion, and D - noisy data with channel distortion.

The ASR  performance on the Aurora-4 dataset is shown in Table \ref{tab:variants_ASR} for various configurations of the proposed approach and Table \ref{tab:multiData_aurora4} for different baseline features.
In order to observe the impact of different components of the proposed model, we tease apart the components and measure the ASR performance (Table \ref{tab:variants_ASR}). The fifth row (A-R,M-R, softmax with no-embedding) refers to the previous attempt using the 2-stage filter learning reported in \cite{agrawal2020interpretable}. In this paper, we explore the variants of the proposed model such as use of softmax nonlinearity instead of sigmoid in both relevance weighting sub-networks, sigmoid in both relevance weighting sub-networks, without and with senone embedding, and the 2-stage approach (both relevance weighting sub-networks). 
Among the variants with acoustic relevance weighting alone, the A-R [sigmoid with senone embeddings] improves over the softmax nonlinearity. With joint A-R,M-R case, again the sigmoid with senone embeddings provides the best result.


While comparing with different baseline features in Table \ref{tab:multiData_aurora4}, it can be observed that most of the noise robust front-ends do not improve over the baseline mel filterbank (MFB) performance. The raw waveform acoustic FB performs similar to MFB baseline features on average while performing better than the baseline for Cond. A and B. The ASR system with MFB-R features, which denote the application of the acoustic FB relevance weighting over the fixed mel filterbank features, also does not yield improvements over the system with baseline MFB features. 
We hypothesize that the learning of the relevance weighting with learnable filters allows more freedom in learning the model compared to learning with fixed mel filters. 
The proposed (A-R,M-R) representation learning (two-stage relevance weighting) provides  considerable improvements in ASR performance over the baseline system with average relative improvements of $15$\% over the baseline MFB features. Furthermore, the improvements in ASR performance are consistently seen across all the noisy test conditions and with a sophisticated RNN-LM. In addition, the performance achieved is also considerably better than the results such as excitation based features (EB) reported by \cite{drugman2015robust}. 

For comparison with the SincNet method by \cite{ravanelli2018interpretable}, our cosine modulated Gaussian filterbank is replaced with the sinc filterbank
as kernels in first convolutional layer (acoustic FB layer in Fig. \ref{fig:block_diag}). The ASR system with sinc FB (Sinc) is trained jointly without any relevance weighting keeping rest of the architecture same as shown in Fig. \ref{fig:block_diag}. From results, it can be observed that the parametric sinc FB (without relevance weighting) performs similar to MFB and also our learned filterbank A. In addition, the relevance weighting with Sinc filterbank (S-R,M-R) results show that the relevance weighting is also applicable to other prior works on learnable front-ends. 
\begin{table}[t]
\centering
\vspace{-0.2cm}
\caption{Word error rate (\%) in CHiME-3 Challenge database for multi-condition training.}
\label{tab:Chime3Results}
  \resizebox{0.9\linewidth}{!}{
	\begin{tabular}{l|c|c|c|c|c|c}
	\hline
		Test Cond & MFB & PFB & {RAS} & {MHE} & A-R & A-R,M-R \\ \hline
		Sim\_dev & 12.9 & 13.3 & 14.7 & 13.0 & {12.4} & \textbf{11.9}\\ 
		Real\_dev & 9.9 & 10.7 & 11.4 & 10.2 & {9.9}  & \textbf{9.5}\\ \hdashline
		Avg. & 11.4 & 12.0 & 13.0 & 11.6  & {11.2}  & \textbf{10.7}\\ \hline \hline

		Sim\_eval & 19.8 & 19.4 & 22.7 &  19.7 & {19.0} &  \textbf{18.7}\\ 
		Real\_eval & 18.3 & 19.2 & 20.5 & 18.5  & {17.2}  & \textbf{17.0}\\\hdashline
		Avg. & 19.1 & 19.3 & 21.6 &  19.1 & {18.1}  & \textbf{17.8}\\ \hline 

	\end{tabular}
	}
\vspace{-0.2cm}
\end{table}

\begin{center}
    \begin{table}[t!]
        \centering
        \caption{WER (\%) for cross-domain ASR experiments.
        }
        \label{tab:cross_domain_ASR}
      \resizebox{0.65\linewidth}{!}
      {
        \begin{tabular}{c|c|c}
        \hline
         {\textbf{Filters  }} & \multicolumn{2}{c}{\textbf{ASR Trained and Tested on}} \\ \cline{2-3}
         \textbf{Learned on} & Aurora-4 & CHiME-3   \\\hline
         Aurora-4 & 9.1 & 14.3 \\
         CHiME-3 & 9.2 & 14.2  \\\hline
        \end{tabular}
        }
        \vspace{-0.3cm}
    \end{table}
\end{center}
\begin{center}
    \begin{figure}[t]
        \centering
        \includegraphics[trim={0 0 0 0cm}, clip, scale=0.32]{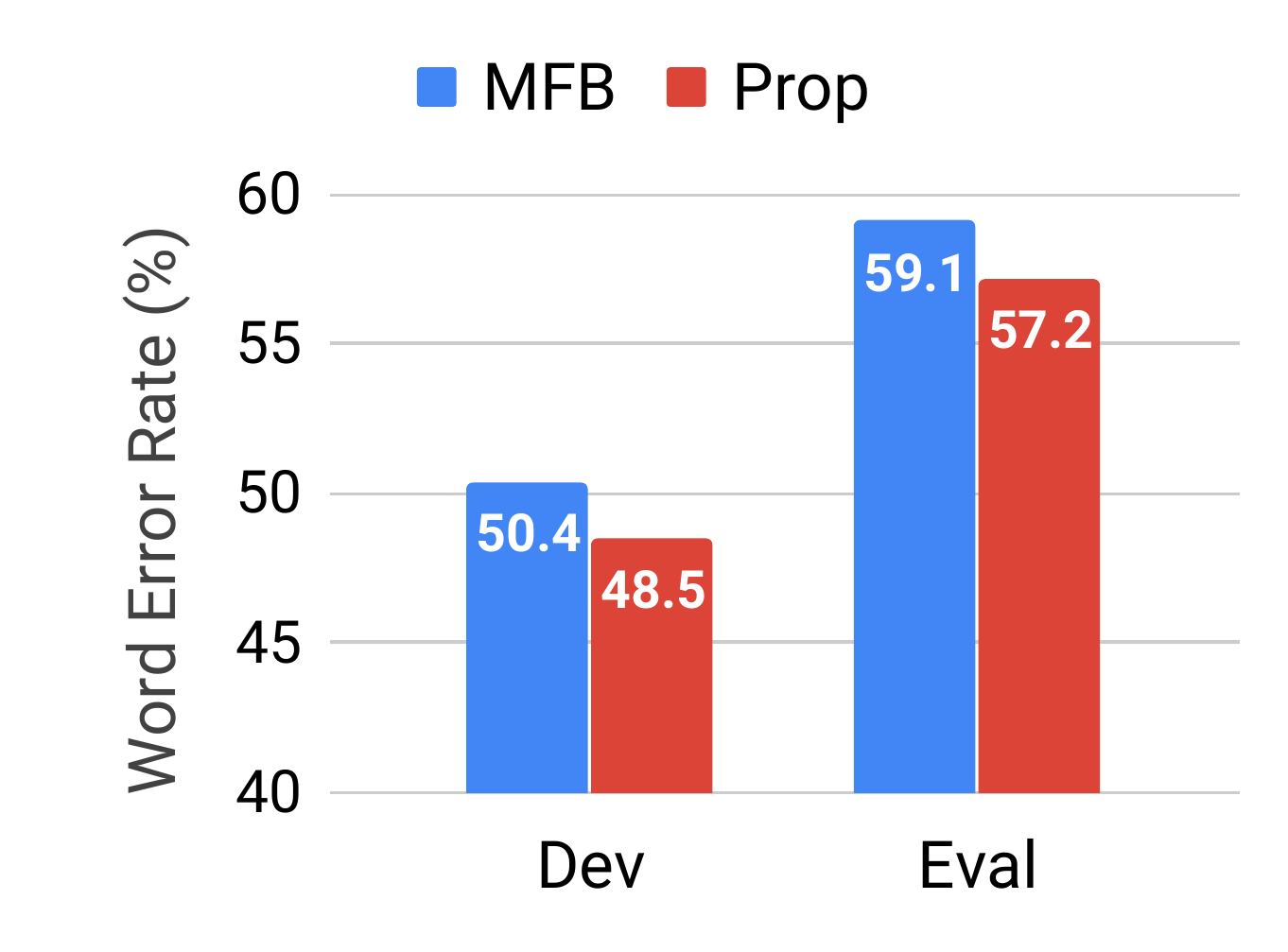}
        \vspace{-0.3cm}
        \caption{ASR performance in WER (\%) for VOiCES database.}
        \label{fig:voices_asr}
        \vspace{-0.3cm}
    \end{figure}
\end{center}
\vspace{-1.6cm}
\subsection{{CHiME-3 ASR}}
The CHiME-3 corpus for ASR contains multi-microphone tablet device recordings from everyday environments, released as a part of 3rd CHiME challenge \cite{barker2015chime3}. Four varied environments are present - cafe (CAF), street junction (STR), public transport (BUS) and pedestrian area (PED). For each environment, two types of noisy speech data are present - real and simulated. The real data consists of $6$-channel recordings of sentences from the WSJ$0$ corpus spoken in the environments listed above. The simulated data was constructed by artificially mixing clean utterances with environment noises. The training data has $1600$ (real) noisy recordings and $7138$ simulated noisy utterances, constituting a total of $18$ hours of training data.
We use the beamformed audio in our ASR training and testing. The development (dev) and evaluation (eval) data consists of $410$ and $330$ utterances respectively. For each set, the sentences are read by four different talkers in the four CHiME-3 environments. This results in $1640$ ($410 \times 4$) and $1320$ ($330 \times 4$) real development and evaluation utterances. 

The results for the CHiME-3 dataset are reported in Table \ref{tab:Chime3Results}. The ASR system with SincNet performs similar to baseline MFB features. The initial  approach of raw waveform filter learning with acoustic FB relevance weighting (A-R) improves over the baseline system as well as the other multiple noise robust front-ends considered here. The proposed approach of 2-stage relevance weighting over learned acoustic and modulation representations provides significant improvements over baseline features   (average relative improvements of $7$\% over MFB features in the eval set). 

\subsection{Representation transfer across tasks}
In a subsequent analysis, we perform a cross-domain ASR experiment, i.e., we use the acoustic filterbank learned from one of the datasets (either Aurora-4 or CHiME-3 challenge)  to train/test ASR on the other dataset. The results of these cross-domain filter learning experiments are reported in Table~\ref{tab:cross_domain_ASR}.
The rows in the table show the database used to learn the acoustic FB and the columns show the dataset used to train and test the ASR (all other  layers in Figure~\ref{fig:block_diag} are learned in the ASR task). The performance reported in this table are the average WER on each of the datasets. The results shown in Table~\ref{tab:cross_domain_ASR} illustrate that the filter learning process is relatively robust to the domain of the training data, suggesting that the proposed approach can be generalized for other ``matched'' tasks.

\subsection{VOiCES ASR}
The Voices Obscured in Complex Environmental Settings (VOiCES) corpus is a creative commons speech dataset being used as part of VOiCES Challenge \cite{nandwana2019voices}.
The training data set of $80$ hours has $22,741$ utterances sampled at $16$kHz from $202$ speakers, with each utterance having $~12-15$s segments of read speech. 
We performed a 1-fold reverberation and noise augmentation of the data using Kaldi \cite{povey2011kaldi}.
The ASR development set consists of $20$ hours of  distant recordings from the $200$  VOiCES dev speakers. It contains recordings from $6$ microphones. The evaluation set  consists of $20$ hours of distant recordings from the $100$ VOiCES eval speakers and contains recordings from $10$ microphones. The ASR performance on VOiCES dataset with baseline MFB features and our proposed approach (A-R,M-R) of 2-step relevance weighting is reported in Figure \ref{fig:voices_asr}. 
These results suggest that the proposed model is also scalable to relatively larger ASR tasks where consistent improvements can be obtained with the proposed approach.

\section{Summary}
The summary of the work is as follows.
\begin{itemize}
\item Extending the previous efforts in 2-stage relevance weighting approach with the use of embeddings feedback from past prediction.
\item Incorporating the feedback in the form of word2vec style senone embedding for the task of learning representations.
\item Performance gains in terms of word error rates for multiple ASR tasks. 

\end{itemize}


\bibliographystyle{IEEEbib}
\bibliography{refer, refs}

\end{document}